# Comparison of two early warning systems for regional flash flood hazard forecasting

Carles Corral, Marc Berenguer, Daniel Sempere-Torres

Center of Applied Research in Hydrometeorology (CRAHI). Universitat Politècnica de Catalunya, Barcelona, Spain.

Laura Poletti, Francesco Silvestro, Nicola Rebora

CIMA Research Foundation, Savona, Italy.

Corresponding author: Carles Corral

carles.corral@crahi.upc.edu

CRAHI-UPC

Module C4 - S1, c/ Jordi Girona, 1-3. 08034 Barcelona

**Abstract**

The anticipation of flash flood events is crucial to issue warnings to mitigate their impact. This work presents a comparison of two early warning systems for real-time flash flood hazard forecasting at regional scale. The two systems are based in a gridded drainage network and they use weather radar precipitation inputs to assess the hazard level in different points of the study area, considering the return period (in years) as the indicator of the flash flood hazard. The essential difference between the systems is that one is a rainfall-based system (ERICHA), using the upstream basin-aggregated rainfall as the variable to determine the hazard level, while the other (Flood-PROOFS) is a system based on a distributed rainfall-runoff model to compute the streamflows at pixel scale. The comparison has been done for three rainfall events in the autumn of 2014 that resulted in severe flooding in the Liguria region (Northwest of Italy). The results obtained by the two systems show many similarities, particularly for larger catchments and for large return periods (extreme floods).

**Keywords**

Flash floods, early warning systems, radar rainfall, drainage network, return period

# 1. Introduction

Flash floods produce devastating effects worldwide. They affect steep and small to medium catchments (up to a few hundreds of square kilometres), and are typically induced by heavy rainfall in the upstream area. The lapse time between strong rainfall (the cause) and the flood (the effect) depends primarily on the features of the catchment, ranging from the order of tens of minutes in the smallest catchments to a few hours in the larger ones (Borga et al., 2008; Gaume et al., 2009; Marchi et al., 2010). In any case, this time is very short, frequently too short to activate the emergency response and effectively prevent damages in human activities, properties and life. Thus, anticipating this kind of hazard and getting some extra minutes or hours for better preparedness and response to mitigate their impact is a crucial and major challenge, especially because of the increase of their occurrence as result of climate change (Munich Re, 2018; Insurance Information Institute, 2017; Barthel and Neumayer, 2012).

Among the difficulties encountered to increase the response time, the most important is the explosive nature of the rainfall events, usually characterized by convective systems producing highly variable rainfall that are very difficult to measure and forecast. Also, flash floods affect catchments at regional scale in areas that are often poorly or very poorly gauged, amplifying the intrinsic difficulties to adjust a rainfall-runoff model. In this context, the high space and time resolution of radar rainfall observations provide relevant information to produce both quantitative precipitation estimates (QPE) and forecasts (QPF), which are fundamental for improving the efficiency of flash flood early warning systems (EWS).

Classically, flood EWS provide real-time information of flood forecasts in specific river sections, which are used to issue warnings to support the response. These warnings are commonly triggered by the exceedance of a particular threshold, for instance the river bank overflow or a statistical frequency of occurrence). However, the limitations to implement an operational rainfall-runoff model over a large region and the difficulties to obtain reliable flow thresholds at the same scale have yield to alternative approaches. And since the rainfall on the upstream basin is the most important forcing factor to produce flow in a given point, this value (spatially averaged over the basin) can be used as the driving variable, leading to systems operationally simpler, although (at least in theory) less accurate. In what follows we will classify flood EWSs in these two main categories, and we will refer to them as flow-based or rainfall-based EWSs (Hapuarachchi et al., 2011).

There have been several attempts over the world to implement flash flood EWSs to operationally produce distributed warnings at regional scale, and not just over gauged river sections (see for example the reviews of Alfieri et al., 2012 and Hapuarachchi et al., 2011). Usually this regional scale corresponds to a political region or country, or to a physical catchment managed by a water authority.

The Flash Flood Guidance concept (FFG) is the base of one of the first operational flash flood EWSs. It consists in running a rainfall-runoff model in inverse mode to determine the amount of rainfall of a given duration needed to produce the overflow of the river bank. It was developed first by the US National Weather Service (NWS, see Clark et al., 2014) and operationally implemented by the Hydrological Research Center (HRC). It was first applied in Central America in 2004 as CAFFG, comparing the rainfall estimated in real time with the FFG values. CAFFG uses satellite rainfall estimates, and the FFG thresholds were calculated using GIS and land surface properties (Georgakakos, 2005; Georgakakos, 2006; Modrick et al., 2014). In the USA, the Weather Forecast Offices of the NWS operate since 2010 this FFG scheme for regional flash flood forecasting in small streams, typically with a lapse time of less than 6 hours (Clark et al., 2014). The FFG system has also been extended to several regions in the framework of a WMO Programme (Georgakakos, 2018).

At European scale, the European Flood Awareness System ( EFAS; Thielen et al. 2009, www.efas.eu) has been the first operational flood EWS. Initially, it was developed to increase preparedness for riverine floods in the EU using the hydrological model LISFLOOD, operated at 5x5 $km^2$ resolution using COSMO-LEPS mesoscale ensembles as rainfall forecasts.

Another operational flash flood EWS, the FF-EWS, was developed by CRAHI as part of the EHIMI project (2001-2010), and operationally applied in 2008 at the Water Agency of Catalonia (ACA) in Spain (Corral et al., 2009). The FF-EWS uses for the first time the basin-aggregated rainfall over the drainage network (explained in detail in section 2) to derive flash flood warnings over the Catalonia region using radar-based QPEs and QPFs.

The concept of the rainfall-based FF-EWS was further developed in the framework of the FP7 EC project IMPRINTS (Sempere-Torres et al., 2011; Alfieri et al., 2011) giving raise to the European Precipitation Index based on Climatology (EPIC; Alfieri and Thielen, 2015) integrated operationally in EFAS since 2012. The EPIC index covers Europe and uses NWP forecasts of basin-aggregated rainfall for flash flood forecasting at a resolution of 1x1 $km^2$. The hazard is computed comparing these basin-aggregated forecasts with a 20-year climatology of the same variable obtained from COSMO reforecasts. Since then, the EPIC index has evolved and has been recently replaced by ERIC (Raynaud et al., 2015), which takes into account the antecedent soil moisture.

The FF-EWS system has been recently implemented at European scale in the context of the DG-ECHO project ERICHA (www.ericha.eu; Park et al., 2017; 2019), and it runs operationally as the radar-based flash flood forecasting module of the European Flood

Awareness System, EFAS since March of 2017. The ERICHA system uses the radar composites produced by the EUMETNET programme OPERA to generate flash flood forecasts at European scale with a resolution of 1x1 km$^2$ in real time. In this way, the ERICHA forecasts (obtained with radar QPE and QPF) become complementary to the ERIC product that uses NWP rainfall forecasts).

In France, SCHAPI operates the flow-based AIGA system (Javelle et al., 2017). It uses a distributed rainfall-runoff model of the GR family to obtain flow forecasts over a gridded drainage network at 1 h and 1x1 km$^2$ resolution. The warnings are provided in the form of return period exceedances using regionalized peak flow quantiles as thresholds.

In summary, a number of operational flash flood EWS prefer the simplicity of the rainfall-based methods due to their lower implementation cost, data needs and computer requirements. Moreover, these methods have shown clear capacity to detect flood events using rainfall threshold exceedances.

This capacity was first described by Guillot and Duband (1967) in their GRADEX method. They found that the gradient of the probability distribution of the annual maximum discharge tends to follow asymptotically the probability distribution of the annual maximum daily rainfall for high return periods. This was proved using long historical flood and rainfall records in France to show that the effect of the antecedent soil moisture is more relevant for flood events below 10 years return period, but it becomes less important as the exceedance probability of the rainfall event diminishes. Thus, using the gradient of the probability distribution of the rainfalls to extend the probability distribution of the flows was proved to be a better methodology to estimate high return periods flows than the direct analysis of the flow records. The GRADEX has been used systematically by Electricité de France and was adopted by the International

Committee of Large Dams (ICOLD-CIGB) as one of their reference calculation methods (Duband et al., 1988, CIGB - ICOLD, 1992 and CIGB - ICOLD, 1994). In what regards flash flood EWS, the GRADEX method is a theoretical support to consider that rainfall-based methods can be used successfully to identify flood events with return periods over 10 years.

In this context, the present work aims to perform a comparison between a rainfall-based and a flow-based EWS, comparing their results on selected case studies to understand the effect of the simplifications and testing the hypothesis of their similarity for high return periods.

To carry out this comparison the Liguria region in Northwest Italy has been selected. This region and in general all the Mediterranean coast is an area particularly prone to suffer important flash floods. Large rainfall amounts produced by the combination of the warm sea and the abrupt orography is the main triggering cause, but also steep and narrow streams in catchments where population and urban development have increased in the recent past decades, often without an adequate flood urban planning. The main reason to select this region is the availability of the data required to perform the comparison, including the existence of the flow-based EWS Flood-PROOFS developed by CIMA (Silvestro et al., 2011; Silvestro and Rebora, 2012, Laiolo et al., 2014). It is operating at the Regional Environmental Protection Agency of Liguria (ARPAL), monitoring the entire region. This system is based on the application of a distributed grid-based rainfall-runoff model, and is a good representative of a flow-based flood EWS applied at regional scale.

On the other hand, we have used the rainfall-based FF-EWS developed by CRAHI (Corral et al., 2009) and used in the ERICHA project. The configuration used here has been applied to Liguria and is currently integrated in the real-time Multi-Hazard EWS developed in the H2020 project ANYWHERE, and hereafter it will be referred as the ERICHA system. Both the ERICHA and Flood-PROOFS systems assess the hazard in the form of probability of occurrence (or return period), and the comparison is made in these terms.

We have analysed three important flood events that affected the Liguria region in the autumn of 2014, including the extraordinary flood event of 09 October 2014 (over 100 years return period). For consistency purposes, rainfall inputs are the same for both systems (radar rainfall estimates during the selected events), and rainfall and flow thresholds are based on the same climatological dataset. Although both systems are designed to incorporate rainfall forecasts (QPF), for the sake of clarity only QPE are used in this study, thus concentrating the analysis on the intercomparison.

The paper is structured as follows: Section 2 and section 3 introduce the rainfall-based and the flow-based flood warning systems, respectively. Section 4 describes the study area, and Section 5 deals with the events and rainfall data. Section 6 is about the results. Finally, some concluding remarks are pointed out in Section 7.

## 2. Description of the ERICHA rainfall-based EWS

The ERICHA rainfall-based flood warning system (Corral et al., 2009; Alfieri et al., 2011, 2017; Versini et al., 2014) estimates the probability of occurrence of the observed basin-aggregated rainfall (i.e. the rainfall averaged in the basin upstream of each point of the drainage network) by comparison with the thresholds obtained from a climatology of the same variable.

The basin-aggregated rainfall is computed at each cell of a gridded drainage network considering different rainfall accumulation windows. The particularity of this system is that the accumulation window used to compute the warnings is set equal to the concentration time of the basin. Thus, each cell representing an upstream basin is related to an individual concentration time [currently the Kirpich (1940) formula is used], and computations to obtain the flash flood indicator are made for this duration.

The basin-aggregated rainfall is estimated at each time step for each pixel of the drainage network and is compared to a set of basin-aggregated rainfall thresholds to assign a hazard level in the form of return period. Ideally, these thresholds should be related to a climatology of the basin-aggregated rainfall to assess the probability of exceedance. An interesting method to estimate this climatology is from long series of radar data. However, the lack of such long series to compute statistically significant thresholds, in this study they have been obtained by processing the Intensity-Duration-Frequency (IDF) curves obtained by Boni et. al. (2007). Using rainfall series from 125 raingauges, they applied a regional approach to derive rainfall frequencies over Liguria, choosing the two components extreme value (TCEV) distribution function as the parent distribution of the rainfall growth factor. These curves are spatially distributed over the grid, and they are referred to as point IDF, *IP($D_i$,$F_j$)*. In practise, each rainfall duration and frequency are related to a rainfall intensity map. Subsequently, the *IP($D_i$,$F_j$)* maps are integrated over the drainage network to obtain basin-aggregated IDF maps, referred to as *IC($D_i$,$F_j$)*. Since point precipitation frequency estimates are representative only for a limited area, approximating the areal estimates by averaging corresponding point precipitation frequency estimates will result in overestimation (Pavlovic et. al., 2016). Thus, a simple areal reduction factor is applied when calculating *IC($D_i$,$F_j$)* from point *IP($D_i$,$F_j$)*, given by $K_a$=1-$\log_{10}$(*S*/15) (which is used in Spain for the design of discharge structures, MOPU, 1990), where *S* is the basin area in km$^2$.

If an important reservoir exists upstream of a cell, then the basin is split at the dam level and the upstream basin is not included for basin-aggregated computations (thus assuming total regulation).

Then, given the QPE (and the QPF if available) at a specific operation time, the basin-aggregated rainfall is computed at each cell over the drainage network, considering an accumulation window equal to the basin concentration time (operationally, the system only works with previously specified durations, and it uses the duration value closest to the concentration time). Comparing this against basin-aggregated rainfall thresholds for this specified duration, the frequency (representing the probability to exceed a given value) is derived over the forecasting window, normally expressed as a return period (average time between two occurrences).

This procedure is simple and easy to implement at regional scale, needing only information about rainfall climatology to build the IDF curves, and a terrain elevation DEM to derive the drainage network and the concentration time map.

## 3. Description of the Flood-PROOFS flow-based EWS

The Flood-PROOFS system existing in the Liguria region is based on the application of the Continuum rainfall-runoff model (Silvestro et al., 2013; 2015), a continuous distributed hydrological model based on a simplified gridded drainage network that relies on a morphological approach (Giannoni et al., 2005). It is founded on a physically based description, but a number of simplifications allows the model to keep its parameters in a suitable number for calibration. This system uses a gridded drainage network (for coherence, the same as the ERICHA system). Infiltration and subsurface flows are described using a semi-empirical, but quite detailed, methodology based on a modification of the Horton algorithm (Bauer, 1974; Diskin and Nazimov, 1994;

Gabellani et al., 2008); it accounts for soil moisture evolution even in conditions of intermittent and low-intensity rainfall (lower than the infiltration capacity of the soil). The energy balance is based on the so called ''force restore equation'' (Dickinson, 1988) which balances forcing and restoring terms, with explicit soil surface temperature prognostic computation. Vegetation interception is schematized with a storage which has a retention capacity ($S_v$) estimated with the Leaf Area Index data, while the water table and the deep flow are modelled with a distributed linear reservoir schematization and a simplified Darcy equation.

The surface flow schematization distinguishes between hillslope and channelled flow by means of a morphologic filter defined by the expression $A\,S^k=C$, where $A$ is the contributing area upstream of each cell, $S$ is the local slope, and $k$ and $C$ are constants that describe the geomorphology of the environment (Giannoni et al., 2000). The overland flow (hillslopes) is described by a linear reservoir scheme, while for the channelled flow the kinematic wave approach is applied (Wooding, 1965; Todini and Ciarapica, 2001).

Continuum model has six parameters that require calibration at the basin scale: two for the surface flow routing: hillslope flow motion coefficient ($u_h$) and channel friction coefficient ($u_c$); two for the subsurface flow generation: mean field capacity ($c_t$) and infiltration capacity at saturation ($c_f$); and two for the deep flow and water table components: maximum capacity of the deep aquifer ($VW_{max}$) and the parameter of anisotropy ($R_f$). At the regional scale of Liguria, the model was calibrated using observed flow data from 11 level gauge stations (Davolio et. al., 2017), by means of a multi-objective function based on the Nash-Sutcliffe Efficiency and the relative error on high flows. The six model parameters are assumed to be uniform at basin scale, and average parameters obtained in calibrated basins are imposed where no observations are available.

At each cell of the region, the flow simulations from the Continuum model are compared with a set of flow thresholds to determine the return period. Over the region of Liguria, these thresholds were obtained from a statistical regional analysis (Boni et. al., 2007), where flood peaks were obtained applying a simple and robust rainfall-runoff model to each cell of the drainage network, using rainfall hyetographs given by a particular Alternating Block Method (Chow et. al., 1988) which maximise the peak response. These hyetographs were built based on the same rainfall frequencies (IDF information) that are used as thresholds in the ERICHA system. Boni et. al. (2007) verified that the peak flow annual maxima provided by this method are consistent with the regional flow analysis made using the historical data from 33 stream flow stations.

## 4. Study area

The Italian region of Liguria is a narrow strip of land about 250 km long and 20–30 km wide with an area of about 5400 km$^2$. Because of the mountainous characteristics of the region, the main urban areas and towns have been established along the coast, often at the mouth of a river. Many basins are small, with an area of less than 100 km$^2$, and prone to flash floods. Only a few catchments have a drainage area over 200 km$^2$, but even in these cases their response times are also very short.

The Liguria region has a real-time hydrometeorological network that provides a detailed set of meteorological variables. There are about 150 automatic weather stations, each one equipped with a rain gauge. The region is also covered by a Doppler polarimetric C-band radar, located on Mount Settepani at an altitude of 1386 amsl (see a more detailed description of the network in Silvestro et al., 2016). The rainfall data used in the present study corresponds to QPEs obtained from the observations of the Settepani radar. These rainfall estimates have been produced at 1x1 km$^2$ and 10 minutes resolution, applying an algorithm that selects the optimal relationship between rainrate and dual polarisation radar variables based on a decision tree flowchart (Silvestro et al., 2009). The same QPEs have been used as inputs to both ERICHA and Flood-PROOFS. Both EWS have used the information of a Digital Elevation Model (DEM) of the region with a resolution of 0.005 degrees both in longitude and latitude (at this latitude this means around 400 m in the WE direction and around 550 m in the SN direction). Flow directions were identified based on maximum local slope following the D8 approach (each cell is given a unique direction from its 8 neighbours, O'Callaghan and Mark, 1984), and the drainage network is defined at the same DEM resolution.

Figure 1 shows the DEM over the working domain where the flood warning systems have been applied, covering completely the Liguria region. A sub-area around Genova covering the zones most affected by the analysed events has been selected for displaying the results. The most important streams located in this sub-area are: the Bisagno creek (crossing the city of Genova), the Polcevera creek (with its outlet at the Genova port), the Entella river (formed by the junction between the Lavagna and the Sturla rivers at Carasco), and the Orba and the Scrivia rivers (these two flowing to the North to the Po plain). Six control points have been selected in these streams, allowing to analyse the temporal evolution of the results at local scale. Three of these selected points correspond to the location of existing stream gauges: Genova in the Bisagno creek (Passerella Firpo), Tiglieto in the Orba river and Panesi in the Entella river (see bottom panel of Fig. 1).

## 5. Analysed events

The autumn of 2014 was particularly rainy in Liguria. A concatenation of rainfall events affected the region causing important flash floods in several locations. We have selected three events that produced several local floods: 09 October, and 10 and 15 November.

The event of October was particularly devastating and was in-depth described in a paper concerning the flood of the Bisagno creek (Silvestro et al., 2016). Daily accumulations over 200 mm were observed in a large area between Genova and La Spezia (about 80 km South-East of Genova). Rainfall intensities were very extreme (for example, a raingauge located near Genova recorded more than 130 mm in one hour). The observed peak flow in the Bisagno creek (91 km$^2$) was around 1100 m$^3$/s, and about 800 m$^3$/s in the Entella river (371 km$^2$). Flash floods caused major damages over the region, there was one fatality and material damages were estimated in around 100 Million Euros.

Compared to the event of 09 October, the other two selected events are of lesser importance, and the spatial affection of flooding was also smaller. The event of 10 November mainly affected the Entella river (peak flow about 1300 m$^3$/s), and small overflows were reported in the village of Carasco. On the event of 15 November, the most affected zone corresponds to the coast in the West of Genova, being reported local overflowing in the Polcevera creek around Pontedecimo, and also local damages in other creeks nearby. The Orba river also experienced a significant increase in discharge (peak flow about 500 m$^3$/s at the stream gauge of Tiglieto), but without overflowing the main channel. A summary of the effects produced in these three events can be found in https://it.wikipedia.org/wiki/Lista_di_alluvioni_e_inondazioni_in_Italia, and with less detail in http://floodlist.com/europe/1-killed-genoa-floods-italy.

As a way to summarize the quality of the radar-based QPEs used in the study, Figure 2 shows a comparison between raingauge observations and the collocated radar estimates in terms of both hourly and total rainfall accumulations (two scatter plots for each event). For the comparison, only raingauges located within 80 km from the radar location have been selected.

The regression line for the event of 09 October 2014 shows a very small bias. In contrast, the two events of November present more scatter and show a clear trend to underestimation (regression slope about 0.70). The apparent differences between the first event and the other two can partly be explained by the spatial extent of the rainfall event, that in the first case was very limited and most of raingauges registered less than 10 mm.

These results show that these radar QPEs are already affected by several error sources, particularly range dependent (due to the Vertical Profile of Reflectivity, beam broadening or path attenuation, for example) and also related to the complex orography of the region (ground clutter, beam blocking). The quality of these products might affect the performance of the two flood warning systems and they may lead to underestimate the warning levels. In the ARPAL operation the radar QPEs are merged with raingauge data in order to minimise this problem. In this study, focused on the intercomparison, we assume that radar-only QPEs have enough quality to provide a valid basis to evaluate differences and similarities of the EWS simulations.

## 6. Results

The results obtained applying the two flood EWS are presented here. The first sections (6.1-6.3) focus separately on each event, summarizing the results on a map of the selected sub-area and showing the results obtained in some control points. Section 6.4 presents the comparison between the two systems globally for the entire region, taking into account all the points of the drainage network and analysing the impact of several factors.

Although both EWS provide results over the entire domain, the results are only displayed and analysed for the pixels with an associated drainage area larger than 4 $km^2$. This threshold is chosen somehow arbitrarily, for aesthetic reasons (mapping is clearer) but also functional because we consider that the estimates obtained in the smallest basins are subject to additional uncertainty. For instance, stream flow gauge stations are located usually in the main streams, and the hydrological model calibration has been done at scales that are far from the smallest basins. Also, at the applied resolution (about 500 m), the geometry of these smallest basins may be badly defined by the simplified drainage network. In these cases, a more detailed description would be needed, which is out of the regional scope of the two analysed flood warning systems.

## 6.1. Event of 09 October 2014

Figure 3 summarizes the results obtained for the Event of 09 October 2014, which is by far the most important of the three cases. The figure shows the maximum return period obtained over the 48-hour time window by each flood warning system (for the pixels with a drainage area over 4 $km^2$). The rainfall accumulation map over the entire region is also included.

Although there was significant rainfall only in part of the region, the 48-hour rainfall accumulation exceeds 200 mm over a very large area, and even 600 mm are exceeded over an area at the north-east of Genova. The maximum accumulation was observed by a raingauge located near Montoggio with a total of 476 mm.

The two systems identified the maximum return periods in the Bisagno creek, the Entella river and the Scrivia river (this last one flowing to the north). Comparing the results provided by both systems, the return periods obtained by ERICHA are in general smaller. Very high return periods were obtained by Flood-PROOFS in the medium course of the Bisagno creek (over 200 years), where the differences with ERICHA (which shows return periods below 100 years) are more evident.

The results of the event are also presented in terms of the time series of estimated return period obtained at 10-min temporal resolution in the most interesting control points (shown with thicker circles in Fig. 3). In addition, when a stream gauge exists at the place, the warning level associated with the observed flow discharge (1 hour resolution) is also included, using the transformation to return period that corresponds to that cell of the drainage network.

Figures 4-6 show the estimated return periods obtained in three control points. The two systems obtained very similar results in the Genova control point (Fig. 4), in particular the peak (both around 125 years), and they provide a very accurate agreement with the return period estimated from flow observations. The coarse time resolution of the flow observations (1 hour) does not allow any conclusion about the performance of the systems during the rising limb (with 10-min resolution). In the case of the Entella river at Panesi (Fig. 5), the warning level provided by Flood-PROOFS (63 years) is quite higher than that of the observed flow (5 years), and ERICHA (7 years) provides a more reliable estimate. Finally, high warning levels are also obtained in the control point of Montoggio in the Scrivia river (Fig. 6), being the return period of Flood-PROOFS (124 years) higher than that of ERICHA (92 years). Results obtained in Genova and Montoggio (return period around 100 years) seem in accordance with the overbank inundations actually reported in these places.

## 6.2. Event of 10 November 2014

The rainfall accumulation map for this event shows the larger amounts distributed around the coast, with two maxima over 200 mm (Figure 7). But only in the Entella river and its tributary Sturla both systems issued significant warnings. Visual comparison of the results provided by the systems shows that Flood-PROOFS provides in general higher return periods, with maxima over 30 years.

Figures 8 and 9 show the results obtained in the control points around the Entella river. Some differences arise between the results obtained in the two control points. In Panesi (on the Entella river with an area of 371 km$^2$) the Flood-PROOFS provides hazard levels over 10 years (16), which is quite similar to that provided by the observed flow (about 13 years), while the maximum provided by ERICHA is 8 years. But in the case of the Carasco control point (outlet of the Sturla river just before flowing into the Entella, with 132 km$^2$), the results are more similar (particularly the rise and the time to peak), although the ERICHA peak (35 years) is higher than that of Flood-PROOFS (25 years). Actual overflows were reported in the village of Carasco (confluence of the Lavagna and Sturla rivers), while downstream in the Entella river there is no proof of overbank inundations. The return periods obtained by both warning systems are in accordance with these reported inundations.

## 6.3. Event of 15 November 2014

As the other events, the event of 15 November 2014 also shows a rainfall pattern around the coast of Liguria (Figure 10), but in this case concentrated mainly in the western part of the region. The most significant signal has been obtained in the Polcevera creek, with maximum return periods over 100 years estimated using the two warning systems, although the river courses where these maxima are located differ from one system to the other.

In this event, only the warning levels obtained in the control points of Pontedecimo (Polcevera creek) and Tiglieto (Orba river) are interesting (Figures 11 and 12). In the first case, there are two rainfall peaks having a time lapse of about two hours. While Flood-PROOFS reflects these peaks with warning levels over 10 years (22 years in the second peak), ERICHA only provides significant warning levels in the second phase (16 years), showing certain delay with respect to the Flood-PROOFS outputs. It is interesting to note that during this event, the Polcevera creek in Pontedecimo and downstream actually suffered minor overbank flooding, which seems in accordance with the estimated maximum return periods.

In the case of the Tiglieto control point (Fig. 12), the outputs from both systems are quite similar around the peak (19 years for Flood-PROOFS; 16 years for ERICHA). And both systems are quite close to the warning level derived from the observed flow available at this place (17 years).

### 6.4. General comparison analysis

The analysis made at the control points allows us to clearly compare the outputs of the two systems. In general, the results obtained at these specific places show significant agreement between the two systems, particularly in what regards the evolution during the event.

To extend the analysis to the rest of the domain, the maximum return periods obtained over the entire event by each system have been selected. The two resulting return period maps have been displayed in a scatter plot. From the total number of pixels of the domain, only those belonging to the main drainage network (having a drainage area greater than 4 km$^2$) have been selected.

The idea of this analysis is to study whether the direct comparison (scatter plots of Figures 13-15) provides information about the differences and similarities of the two warning systems, and if they can be related to other factors. In particular, the analysis focuses on the basin extension (drainage area); the basin itself (since the rainfall-runoff model integrated into the Flood-PROOFS is calibrated at basin scale); and the initial moisture state of the basin (since this is a key variable to generate surface runoff from rainfall in Flood-PROOFS).

Thus, from the total number of points present in this analysis (6277 points in the main drainage network), we have grouped them according to three criteria: (1) their drainage area, having more or less than 50 km$^2$ (for clarity purposes only two classes are chosen); (2) if they belong or not to one of the five main basins in the selected sub-area of Liguria (Bisagno, Scrivia, Entella, Polcevera and Orba); and (3) the average soil moisture of the upstream basin at the beginning of the event, simulated at cell scale by the Continuum model and expressed as the proportion of the model soil reservoir that is full (referred to as Soil Moisture Index, SMI, with 4 classes).

This last point is more complex than the other two, and requires defining the beginning of the rainfall event. But frequently the rainfall events consist of a concatenation of rainfall periods with high spatial variability, and the definition of the beginning of the event for a large area can be difficult and affected by some subjectivity. Since this analysis is focused on the maxima, we have considered that the period that effectively affects the peak results corresponds to the previous time window with a duration given by the concentration time of the catchment (which is consistent with the background of the rainfall-based warning system). Thus, since peak values are achieved at different times and concentration times vary at each point of the drainage network, each point has been associated with a different state of the soil reservoir simulated by the Continuum model.

Although the results of the three events are included in this section, the events are analysed separately (Figures 13-15). The scatter plot axes show the Gumbel reduced variable $u$, which is related to the return period $T$ (in years) as:

$$u = -ln\left[-ln\left(\frac{T-1}{T}\right)\right] \qquad \text{(Eq. 1)}$$

The upper-right scatter plots of Figures 13-15 do not seem to show any clear relationship with the location (see catchment grouping). In fact, the results obtained in any of the five analysed catchments present a high scatter. In contrast, the explanation of this scatter seems to be mainly due to the points having an associated catchment area of less than 50 km² (upper-left scatter plots). If we keep only the points with a basin area greater than this value, a clear trend is observed: in general, ERICHA warning levels are smaller than those obtained by Flood-PROOFS, particularly for small return periods (almost systematically), while for higher return periods the two warning systems present a better agreement (although the general trend remains to be that ERICHA return periods are a bit smaller). This result is in accordance to the hypothesis highlighted in the introduction that, for small probabilities of occurrence, discharge and rainfall probabilities tend to follow the same gradient (for an analysis based on annual maxima).

The analysis grouping the pixels according to the initial soil moisture of the upstream catchment (Figures 13-15 bottom) shows a logical evolution after these important rainfall events over a period of less than 25 days: while many catchments are relatively dry at the beginning of the first event (October 2014), they are almost completely wet at the beginning of the third event. In any case, the event of October was characterised by a strong rainfall persistence over the affected catchments, and the state of the soil reservoir of these catchments was very wet prior to the main rainfall event (SMI higher than 0.70).

It is interesting to note that, as a general fact, high return periods are not obtained in catchments that are dry at the beginning of the event. There are only a few catchments having a Soil Moisture Index of less than 0.70 where a return period of 8 years is exceeded ($u>2$). In fact, only a few points having an initial SMI smaller than 0.50 appear in the scatter plots (mainly in the event of 09 October, Figure 13 bottom panel), meaning that at least one of the systems did not obtain a significant return period, and thus it is not possible to derive any conclusion for dry conditions. As it has been stated, the general trend is that Flood-PROOFS peaks are higher than ERICHA peaks, and this is apparent for the points having the largest values of SMI (0.70-1.00), and also for medium values of SMI (0.50-0.70).

In order to somewhat quantify the similarities of the two systems, the Critical Success Index (CSI, Schaefer, 1990) between the Flood-PROOFS and ERICHA is computed for the peak values over certain thresholds. Usually, the CSI is computed when comparing an estimate to an observation, but it can also be used to compare two variables, providing a measure of agreement, and the value of the CSI is independent of which is used as reference. The CSI has been evaluated over the whole area, and also following the classification with the drainage area, catchment pertaining and initial soil moisture.

Table 1 summarizes the results for return period thresholds of 2, 10 and 20 years. The 2-years return period could be considered as a proxy for bankfull conditions, and the 10-years event for a significant flooding.

One can observe high CSI values in the catchments that were affected by extreme floods, as it is the case for Bisagno, Entella and Scrivia in the event of 09 October and for Polcevera in the event of 15 November. When CSI is computed for return periods over 10 years, then it is clear that better agreement is obtained in the catchments with a drainage area over 50 km$^2$ than in smaller catchments. And when looking at the influence of the initial soil moisture, better agreement is obtained in the catchments with SMI values higher than 0.70 than in drier catchments (in general it seems to apply for all the different return period thresholds tested), but it is probable that the CSI obtained for SMI values lesser than 0.70 would be contaminated by the fact that there are too few effective points with a related peak over the threshold (for at least one of the warning systems), making the CSI computation very sensitive.

**7. Discussion and conclusions**

The comparison of the two flood EWS in the three selected events show remarkable similarities in the three control points where objective assessment can be performed. This similarity is very clear in the case of the event of 9 October 2014 at the Bisagno river (Genova) for which both systems produce excellent estimates of the observed flows. This agreement with observations supports the quality of the calibration of the Flood-PROOFS EWS, which allows it to reproduce even this exceptional flood event (return period over 100 years). But it becomes outstanding in the case of ERICHA EWS because it only relies on the quality of the rainfall inputs and on the rainfall thresholds to estimate the return period probability level (with no other components to be calibrated). This can be seen as a good example of the consistency of the GRADEX hypothesis for the flood events with high return periods.

Moreover, all the results at the available control points show quite a good agreement between the two EWS in terms of the evolution and magnitude of the estimated hazard level. And, from the comparison with the existing flow observations, it is not clear that the flow-based flood EWS (Flood-PROOFS) performs better than the rainfall-based EWS (ERICHA), specially for high return periods.

The representativeness of this comparison is somewhat limited by the fact that the observed flow data have been only available at 3 stations, and the time series have been compared at 6 selected control points. However, the significance of the selected events allows a relevant analysis since it is not common to perform these assessments in events with return periods over 10 years.

When comparing both EWS over the entire domain (the Liguria region), the results show more differences. Contrasting the maximum return period obtained at each pixel by each one of the systems in a scatter plot, the agreement between Flood-PROOFS and ERICHA can be considered low. As a general trend, return periods provided by ERICHA are lower than those provided by Flood-PROOFS, although the agreement is clearly closer when dealing with extreme floods, particularly for larger catchments (above 50 km$^2$).

It is worth noting that both EWS are affected by the quality of the rainfall inputs. In particular, systematic biases (such as those affecting the events of November 2014, Fig 2) can affect their skill in identifying the areas affected by floods. In this work, the uncertainty due to QPE might affect the outputs of the two EWSs and their comparison against observations (measured at the 3 stream gauge stations), but much less the comparison of the results obtained with the two systems.

The rainfall-based flood EWS explored in this paper has important similarities with the European Precipitation Index EPIC presented by Alfieri et al. (2011). This index is the basin-aggregated rainfall normalised by the mean of annual maxima, computed for different rainfall durations (6, 12 and 24 h) and keeping the value that provides the maximum index. Alfieri and Thielen (2015) made a comparison similar to that shown in Figures 13-15, comparing the EPIC index ($K_{Pmax}$) against a parallel $K_Q$ index obtained from the simulated discharges (obtained with the rainfall-runoff model used in EFAS) normalised by the mean of annual maximum discharges. Their results showed that the rainfall index (EPIC) generally provided higher return periods than the flow index ($K_Q$), over the full range of results (return periods ranging from 1 to 200 years). Thus, these results were notably different than that of the present paper, where the general trend is that ERICHA provides lower return periods in front of Flood-PROOFS.

Raynaud et al. (2015) proposed a sophistication of EPIC to include the effect of the soil moisture in the runoff coefficient, and they found that the new index (ERIC) provided better agreement against observations than EPIC, so inherently showing the importance of including soil moisture information. Contrarily, our results do not show a clear effect of the initial soil moisture conditions on the comparison between the two systems. At least in part, this can be related to the fact that the basins affected by heavy rainfall were quite wet prior to the beginning of the main rainfall event. But it can also be related to the own dynamics of the Continuum model during the rainfall event, particularly that of the runoff generation. These issues should be addressed in the future, extending the comparison to a larger number of events including more varied initial conditions.

Furthermore, the performance of the rainfall-based flood warning system used in this study could be dependent on the duration used to accumulate rainfall. Since this duration is related to the concentration time of the basin, the formulation used to obtain this parameter might be quite critical. In fact, the concentration time responds to a conceptualisation of the basin, and the existent formulations are usually empirical. A universally accepted working definition of this parameter is currently lacking, and available approaches for the estimation of the time of concentration may yield numerical predictions that differ from each other by up to 500% (Grimaldi et al., 2012). In our case, using Flood-PROOFS as a reference, the selection of the Kirpich formula for the concentration time seem to produce better agreement in terms of the warning level peak than other alternatives, particularly in terms of timing. Our impression is that this characteristic time does not have to be strictly referred to the formal definition of the concentration time (the time it takes a drop of water to travel from the most distant point to reach the outlet), but to a time that allows most of the basin to contribute to the hydrograph. In this sense, the Kirpich method, which provides quite short concentration times in comparison with other approaches, has resulted to be a good estimate in the studied domain. It should be pointed out that the application of the method to other areas could require a previous analysis of the concentration time assigned to the basins, and this would be an opportunity to essay methods that better exploit the geomorphological information.

The rainfall thresholds used in the rainfall-based EWS (ERICHA) have been obtained from historical raingauge series existing at a few specific places. When quality and long series are available, good IDF estimates can be obtained at local scales, but several difficulties arise when transforming them into basin-aggregated IDF curves, since this requires: (i) to spatially interpolate IDF values; and (ii) to apply an "ad-hoc" areal reduction factor. A preferred alternative would have been the direct use of long series of radar QPEs (not available for this study) to directly compute the statistics of basin-aggregated rainfall (i.e. basin-aggregated IDF curves; see Panziera, 2016), and thus leading to a more consistent set of thresholds for the ERICHA EWS.

Finally, as in the majority of real cases, in this study we had not enough streamflow observations to quantify the real magnitude of the analysed flood events in terms of return period at the regional scale (just a few streamflow gauges in the entire region). Then, an accurate analysis of the reliability of the compared EWS has been not possible, and therefore it has not been possible to elucidate whether one flood warning system is better than the other. However, this study can be considered as a significant test to check whether a rainfall-based flood warning system can be reliable and useful.

In general, if a distributed rainfall-runoff model is properly calibrated for a region (with a satisfactory index of agreement against flow observations), if flow thresholds are well defined over the whole region and if computer time requirements are not a major constraint, this kind of approaches should be firstly recommended. As the rainfall-runoff model includes the modelling of the different components of the hydrological system, an EWS able to forecast flows should be better to reproduce the timing of the peaks and provide more confidence and more accurate hazard assessment (particularly in minor events). Thus, in these favourable cases, our recommendation is to use a flow-based EWS.

If it is not the case (and unfortunately this is not unusual), the results of this study support the idea that a rainfall-based flash flood system can be almost as reliable as a flow-based system; in particular for basins larger than 50 km$^2$ and when dealing with flood events with return periods over 10 years. This is reinforced by the fact that the major source of uncertainty in a flash flood EWS is in many cases due to rainfall inputs, particularly to rainfall forecasts, and this uncertainty will affect similarly any of these EWS.

**Acknowledgements**

Table 1. Critical Success Index obtained for the peak values over return period higher than 2, 10 and 20 years. For each event, the points of the drainage network used for the computation correspond to the different classification: whole area; drainage area; catchment pertaining; and SMI (initial soil moisture).

| Event | Classification | CSI T>2 years (u>0.37) | CSI T>10 years (u>2.25) | CSI T>20 years (u>2.97) | # points total |
|---|---|---|---|---|---|
| 09/10/2014 | All points (A > 4 km$^2$) | 0.541 | 0.704 | 0.559 | 6277 |
| | A = 4 – 50 km$^2$ | 0.583 | 0.659 | 0.482 | 4645 |
| | A > 50 km$^2$ | 0.462 | 0.791 | 0.677 | 1632 |
| | Bisagno | 1.000 | 0.982 | 0.893 | 56 |
| | Entella | 0.738 | 0.630 | 0.488 | 231 |
| | Orba | - | - | - | 87 |
| | Polcevera | 0.375 | - | - | 91 |
| | Scrivia | 0.771 | 0.931 | 0.694 | 104 |
| | SMI = 0.00 - 0.50 | 0.087 | - | - | 4791 |
| | SMI = 0.50 - 0.70 | 0.131 | 0.000 | 0.000 | 685 |
| | SMI = 0.70 - 0.90 | 0.498 | 0.441 | 0.373 | 478 |
| | SMI = 0.90 - 1.00 | 0.978 | 0.814 | 0.609 | 323 |
| 10/11/2014 | All points (A > 4 km$^2$) | 0.183 | 0.056 | 0.156 | 6277 |
| | A = 4 – 50 km$^2$ | 0.208 | 0.000 | 0.000 | 4645 |
| | A > 50 km$^2$ | 0.118 | 0.128 | 0.625 | 1632 |
| | Bisagno | 0.000 | - | | 56 |
| | Entella | 0.448 | 0.077 | 0.179 | 231 |
| | Orba | 0.000 | - | - | 87 |
| | Polcevera | 0.112 | 0.000 | - | 91 |
| | Scrivia | 0.000 | - | - | 104 |
| | SMI = 0.00 - 0.50 | 0.000 | - | - | 170 |
| | SMI = 0.50 - 0.70 | 0.208 | 0.000 | - | 2826 |
| | SMI = 0.70 - 0.90 | 0.179 | 0.070 | 0.167 | 3159 |
| | SMI = 0.90 - 1.00 | 0.155 | 0.000 | 0.000 | 122 |
| 15/11/2014 | All points (A > 4 km$^2$) | 0.454 | 0.384 | 0.456 | 6277 |
| | A = 4 – 50 km$^2$ | 0.471 | 0.252 | 0.373 | 4645 |
| | A > 50 km$^2$ | 0.382 | 0.971 | 0.706 | 1632 |
| | Bisagno | 0.625 | - | - | 56 |
| | Entella | 0.000 | - | - | 231 |
| | Orba | 0.874 | 0.447 | - | 87 |
| | Polcevera | 0.846 | 0.667 | 0.554 | 91 |
| | Scrivia | 0.139 | 0.000 | 0.000 | 104 |
| | SMI = 0.00 - 0.50 | 0.000 | - | - | 156 |
| | SMI = 0.50 - 0.70 | 0.000 | 0.000 | 0.000 | 3830 |
| | SMI = 0.70 - 0.90 | 0.357 | 0.471 | 0.442 | 1826 |
| | SMI = 0.90 - 1.00 | 0.947 | 0.324 | 0.500 | 465 |

**List of figure captions**

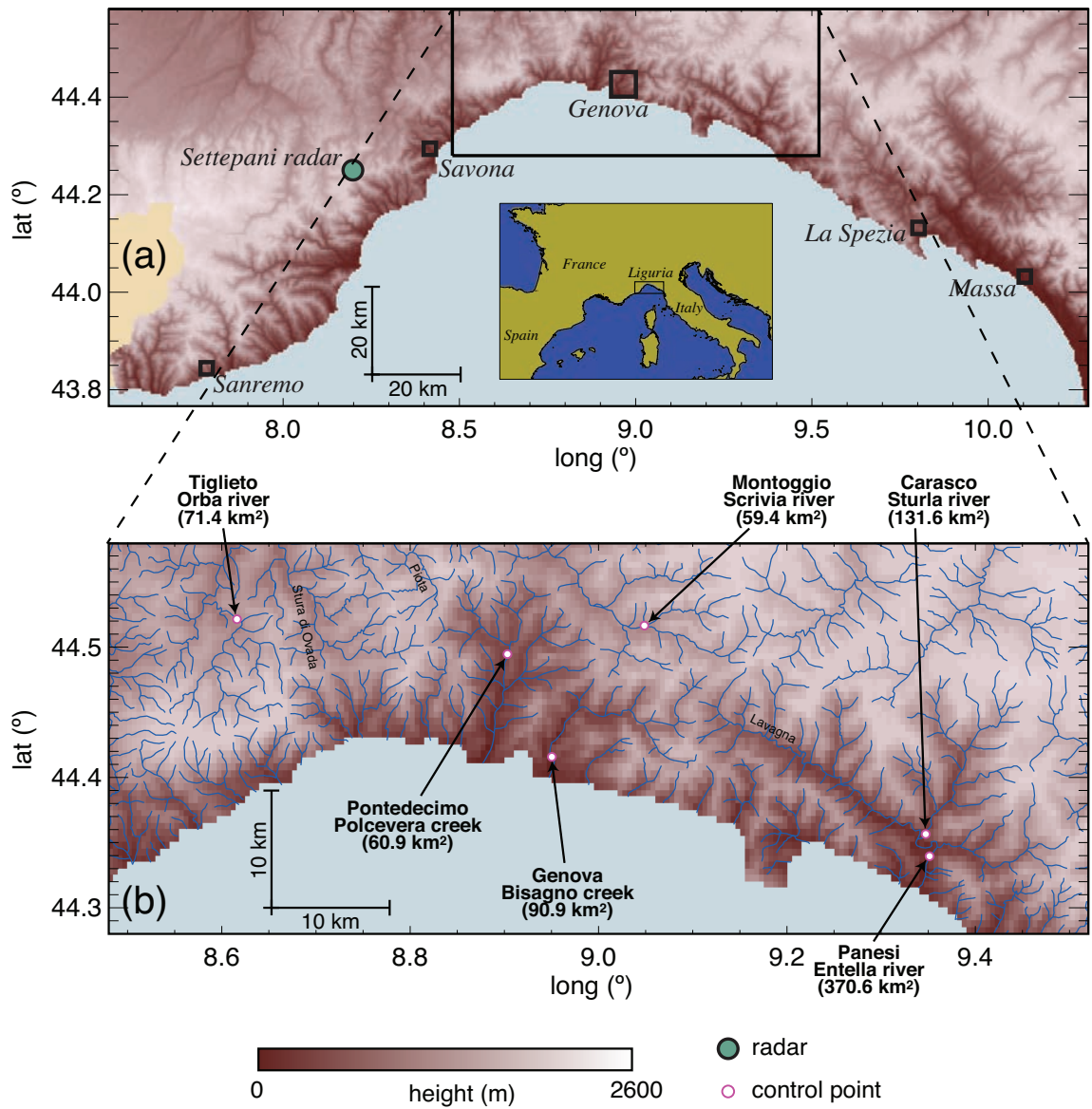

Figure 1. Map of the study area: a) Map from the complete DEM covering the Liguria region and used to apply the warning systems. b) Sub-area where the analysis is focused showing the location of the control points.

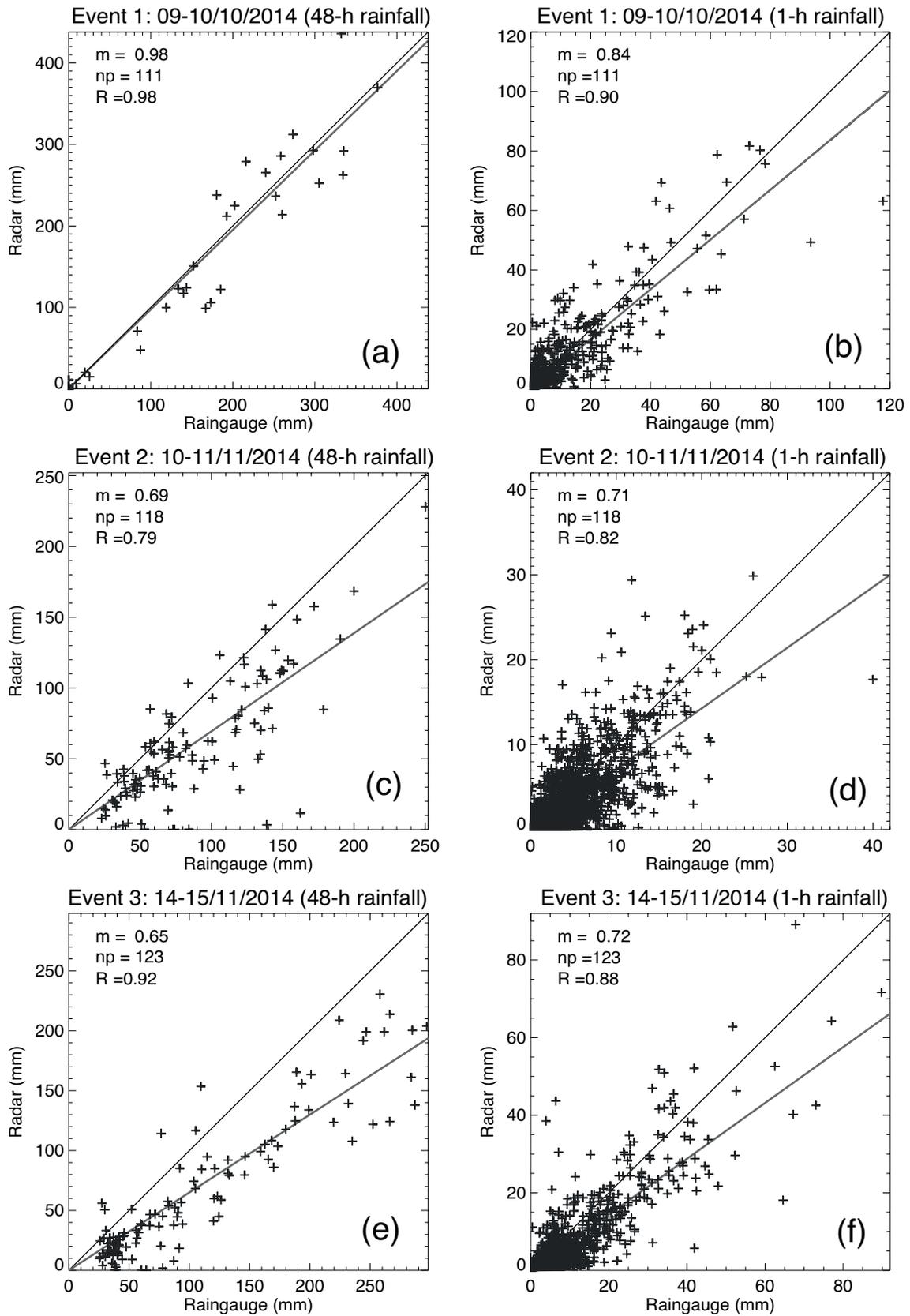

Figure 2. Point radar-raingauge comparison for 1-h and 48-h accumulations (left and right columns, respectively) for the three events: a-b) Event of 09 October: from

09/10/2014 00:00 to 11/10/2014 00:00; c-d) Event of 10 November: from 10/11/2014 00:00 to 12/11/2014 00:00; e-f) Event of 15 November: from 14/11/2014 00:00 to 16/11/2014 00:00. The regression lines and their slope m are included, as well as the Pearson correlation (R) and the number of raingauges (np) included in the analysis inside the 80-km range.

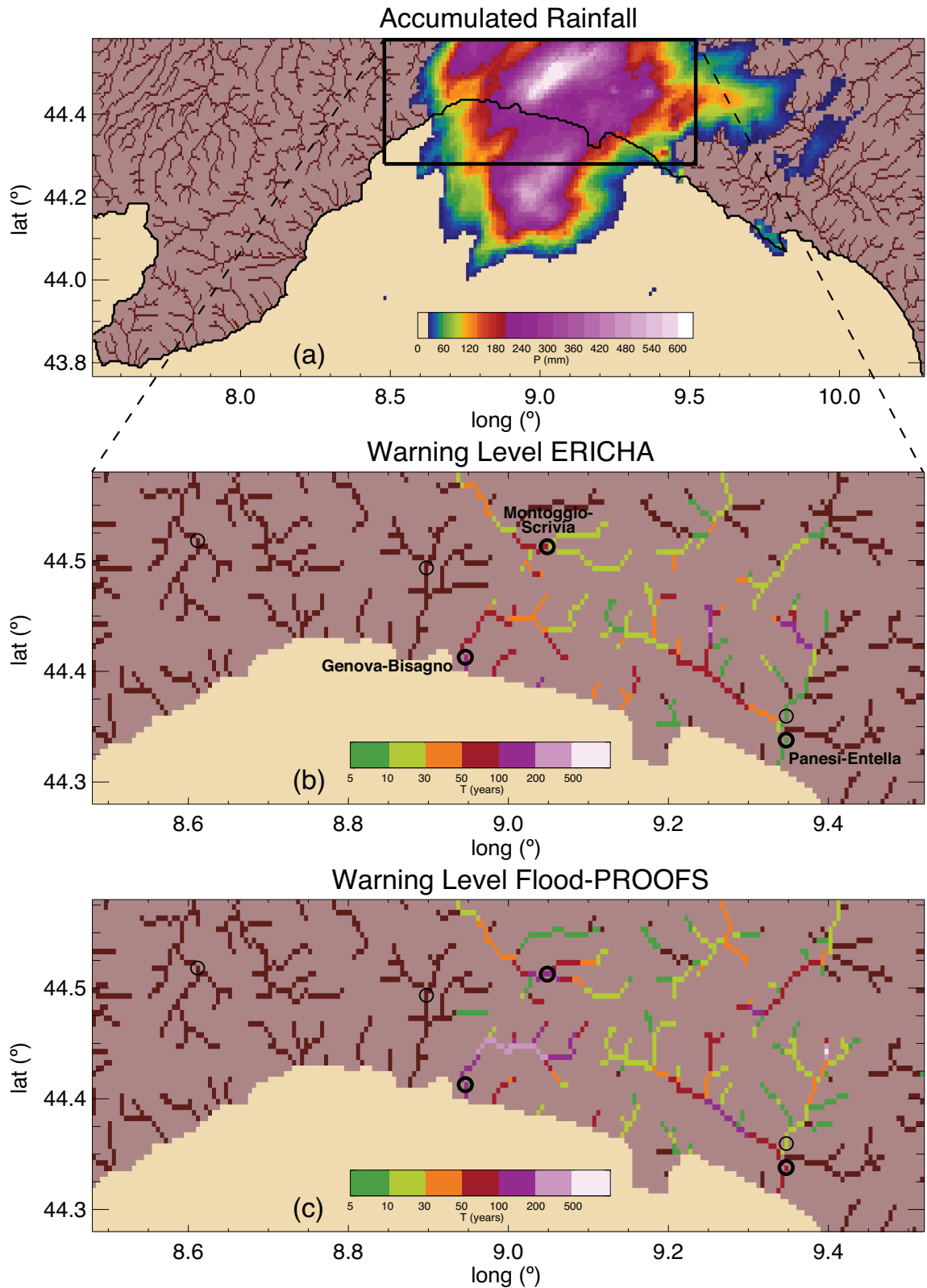

Figure 3. a) Rainfall accumulation for the event of 09 October from 09/10/2014 00:00 to 11/10/2014 00:00 (48-hour accumulation); b) Maximum return period obtained by ERICHA; c) Maximum return period obtained by Flood-PROOFS. Thicker circles

highlight the control points selected to show the temporal evolution in subsequent figures.

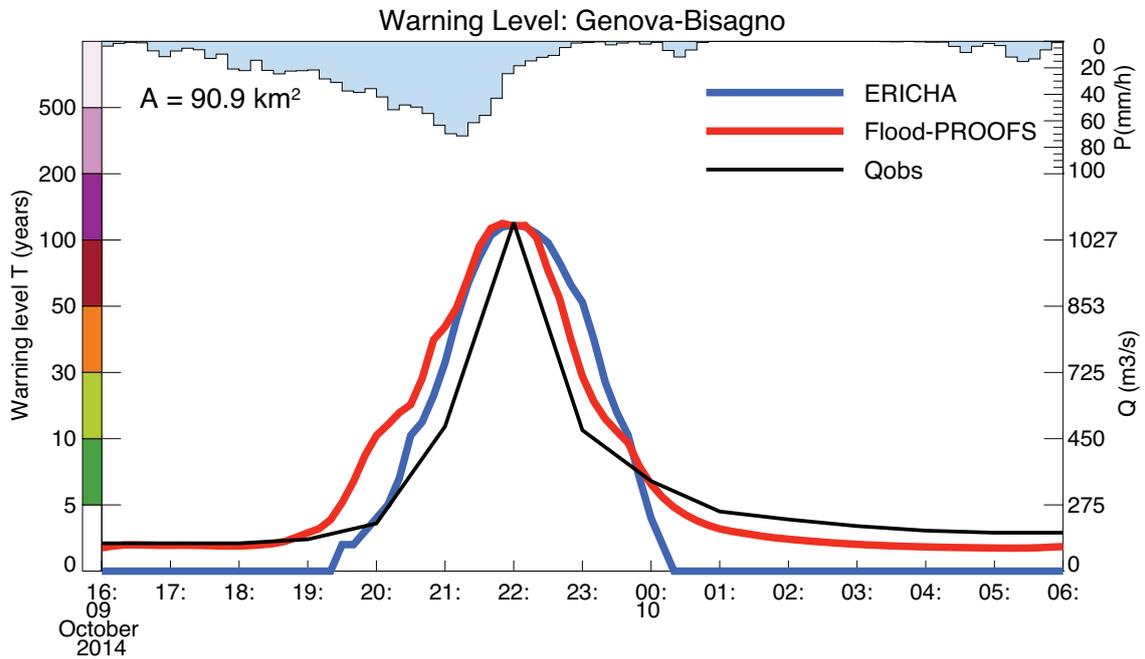

Figure 4. Comparison between ERICHA and Flood-PROOFS outputs for the event of 09 October 2014 at Genova control point (Bisagno creek). Warning level from observed flow is also included (threshold discharge values in the right axis). Upstream catchment rainfall is included.

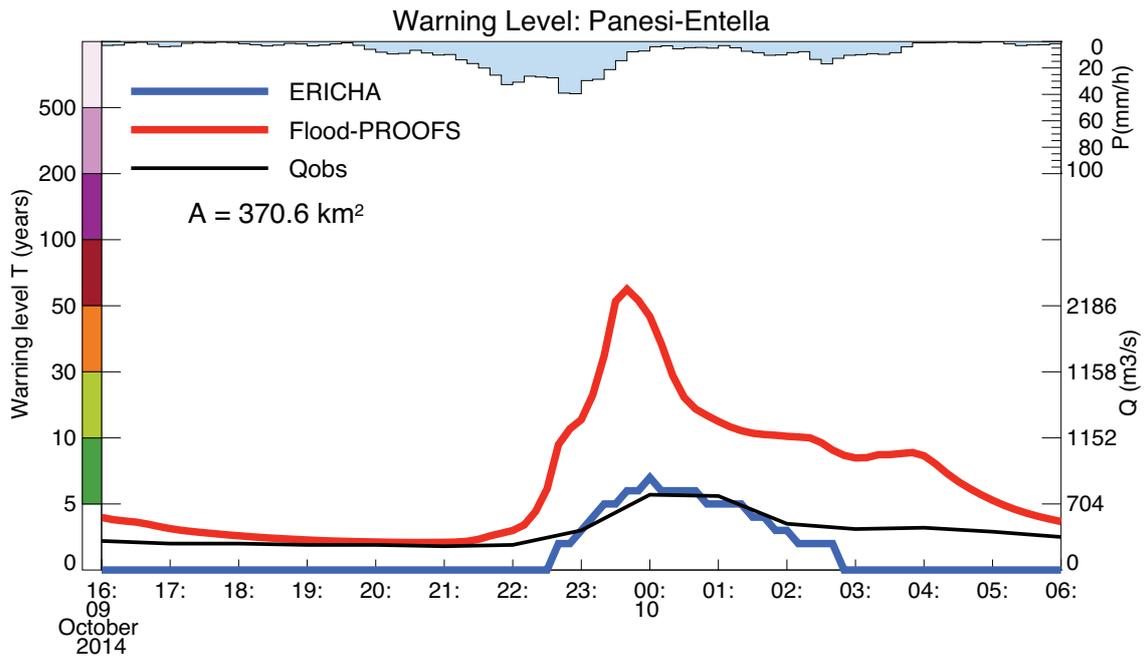

Figure 5. Comparison between ERICHA and Flood-PROOFS outputs for the event of 09 October 2014 at Panesi control point (Entella river). Warning level from observed flow is also included (threshold discharge values in the right axis). Upstream catchment rainfall is included.

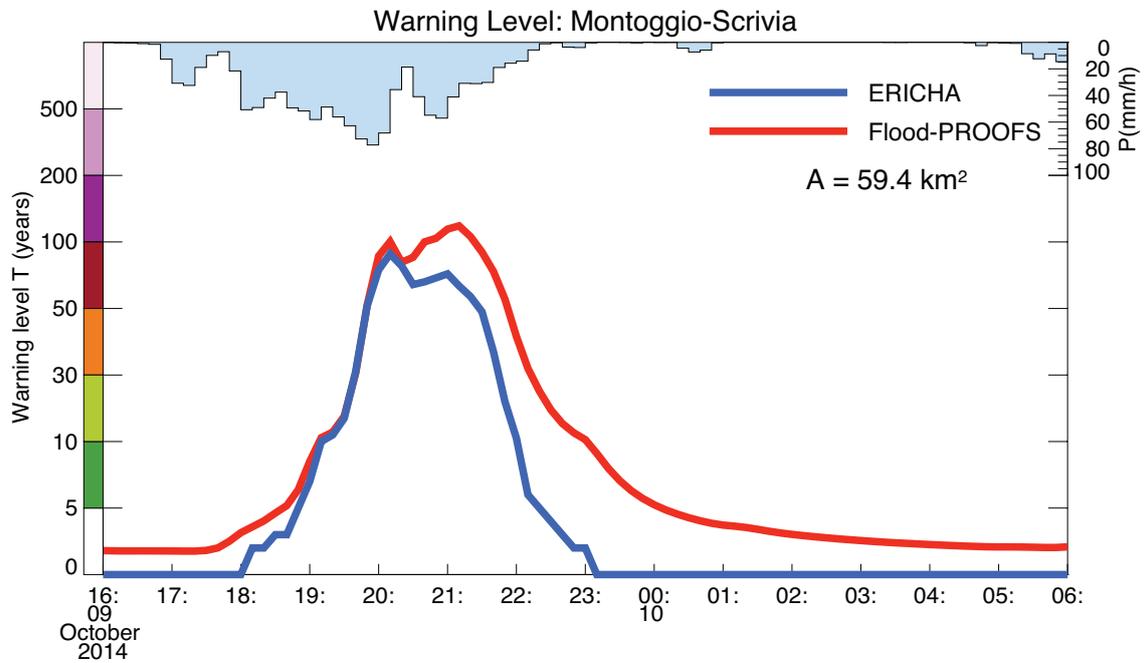

Figure 6. Comparison between ERICHA and Flood-PROOFS outputs for the event of 09 October 2014 at Montoggio control point (Scrivia river). Upstream catchment rainfall is included.

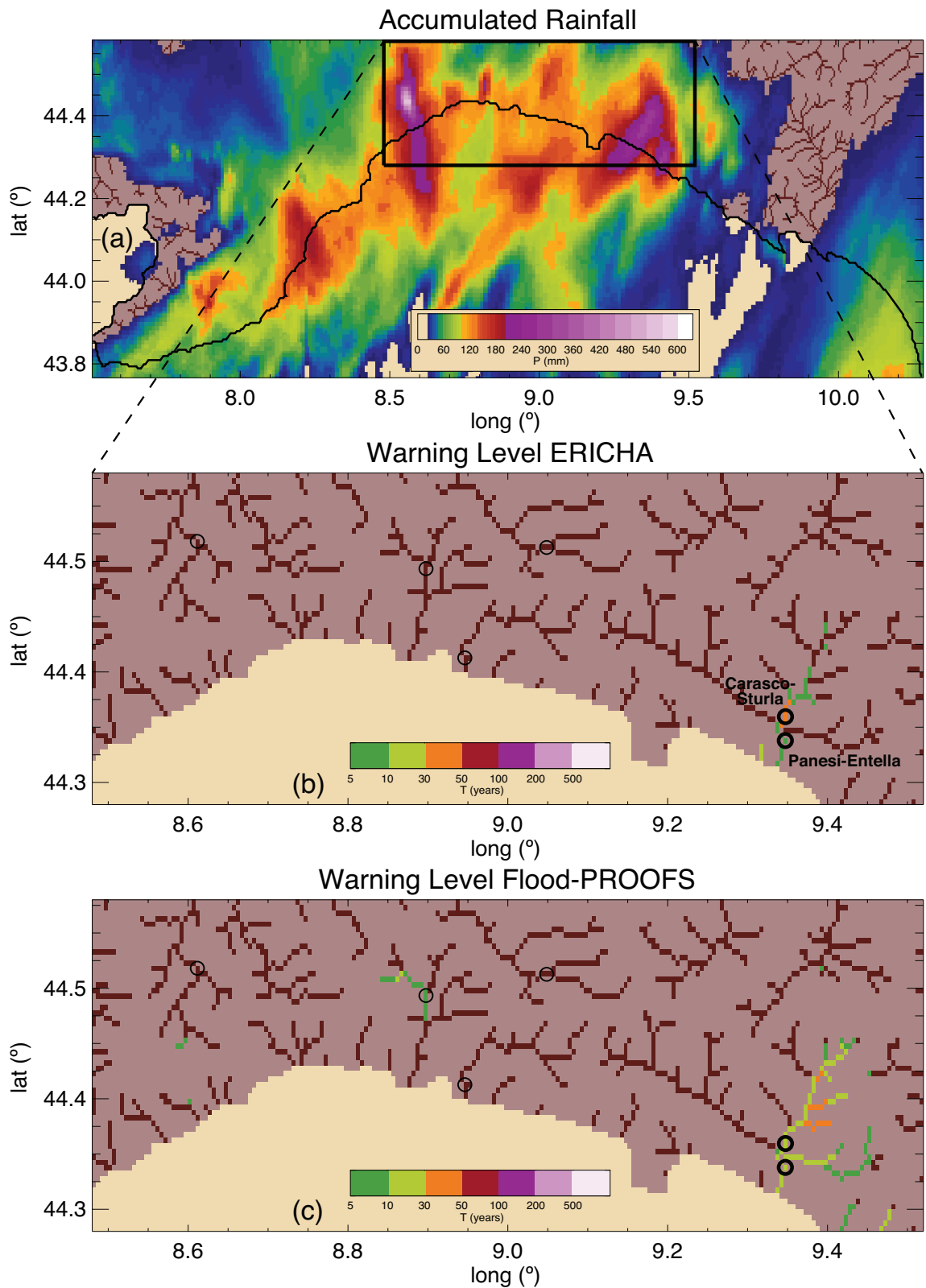

Figure 7. Rainfall accumulation for the event of 10 November from 10/11/2014 00:00 to 12/11/2014 00:00 (48-hour accumulation); b) Maximum return period obtained by ERICHA; c) Maximum return period obtained by Flood-PROOFS. Thicker circles highlight the control points selected to show the temporal evolution in subsequent figures.

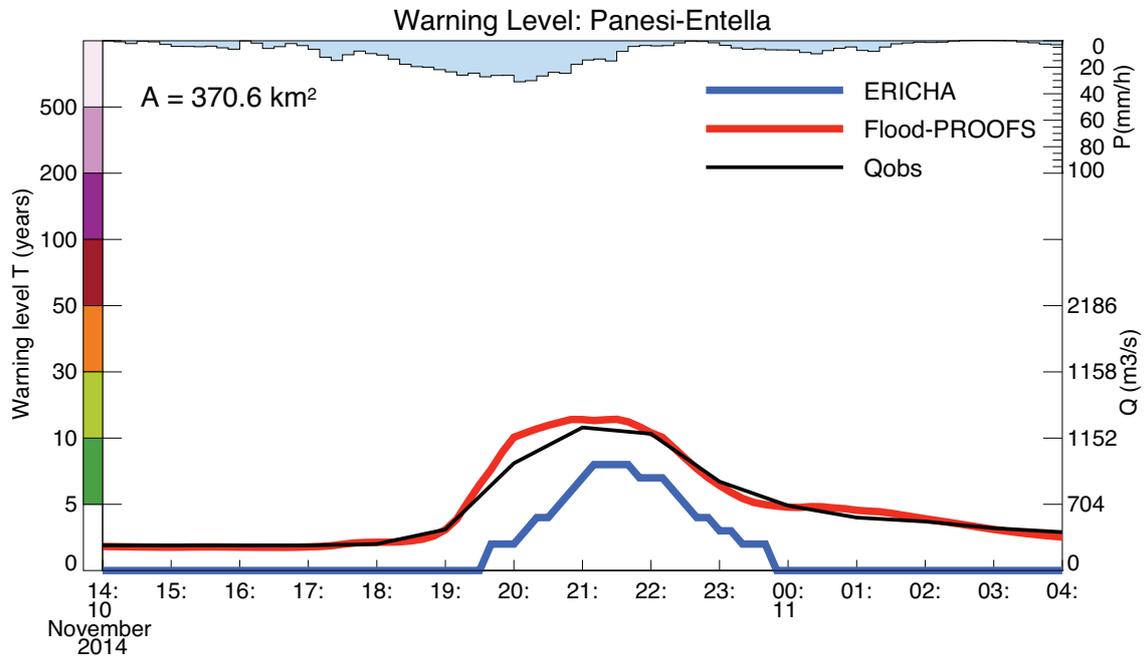

Figure 8. Comparison between ERICHA and Flood-PROOFS outputs for the event of 10 November 2014 at Panesi control point (Entella river). Warning level from observed flow is also included (threshold discharge values in the right axis). Upstream catchment rainfall is included.

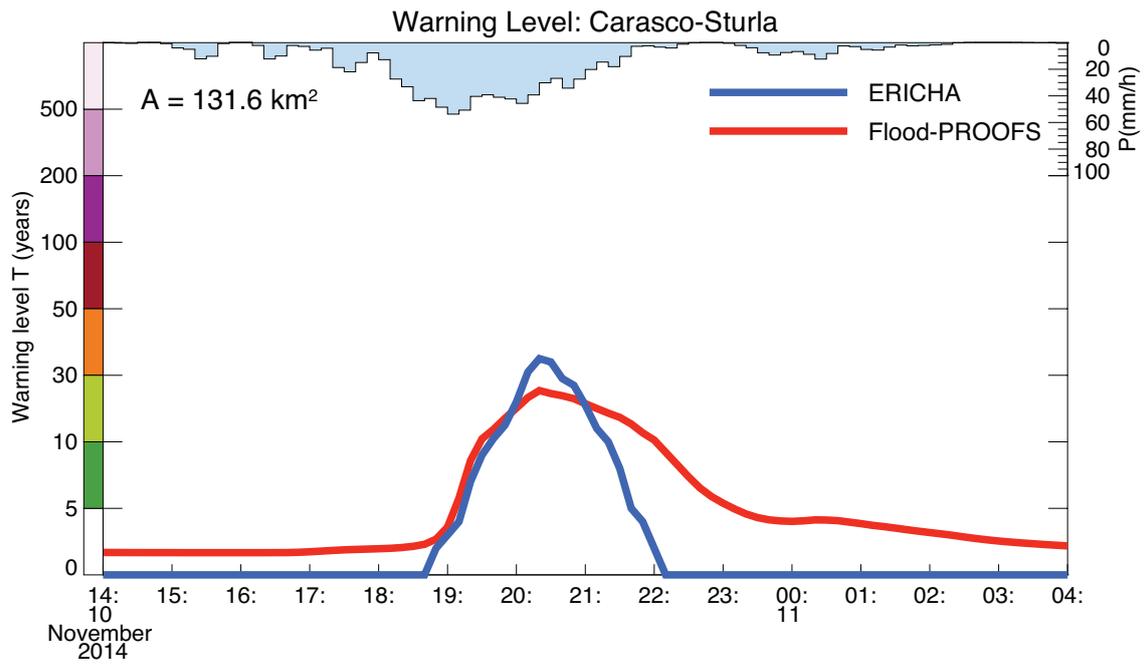

Figure 9. Comparison between ERICHA and Flood-PROOFS outputs for the event of 10 November 2014 at Carasco control point (Sturla river). Upstream catchment rainfall is included.

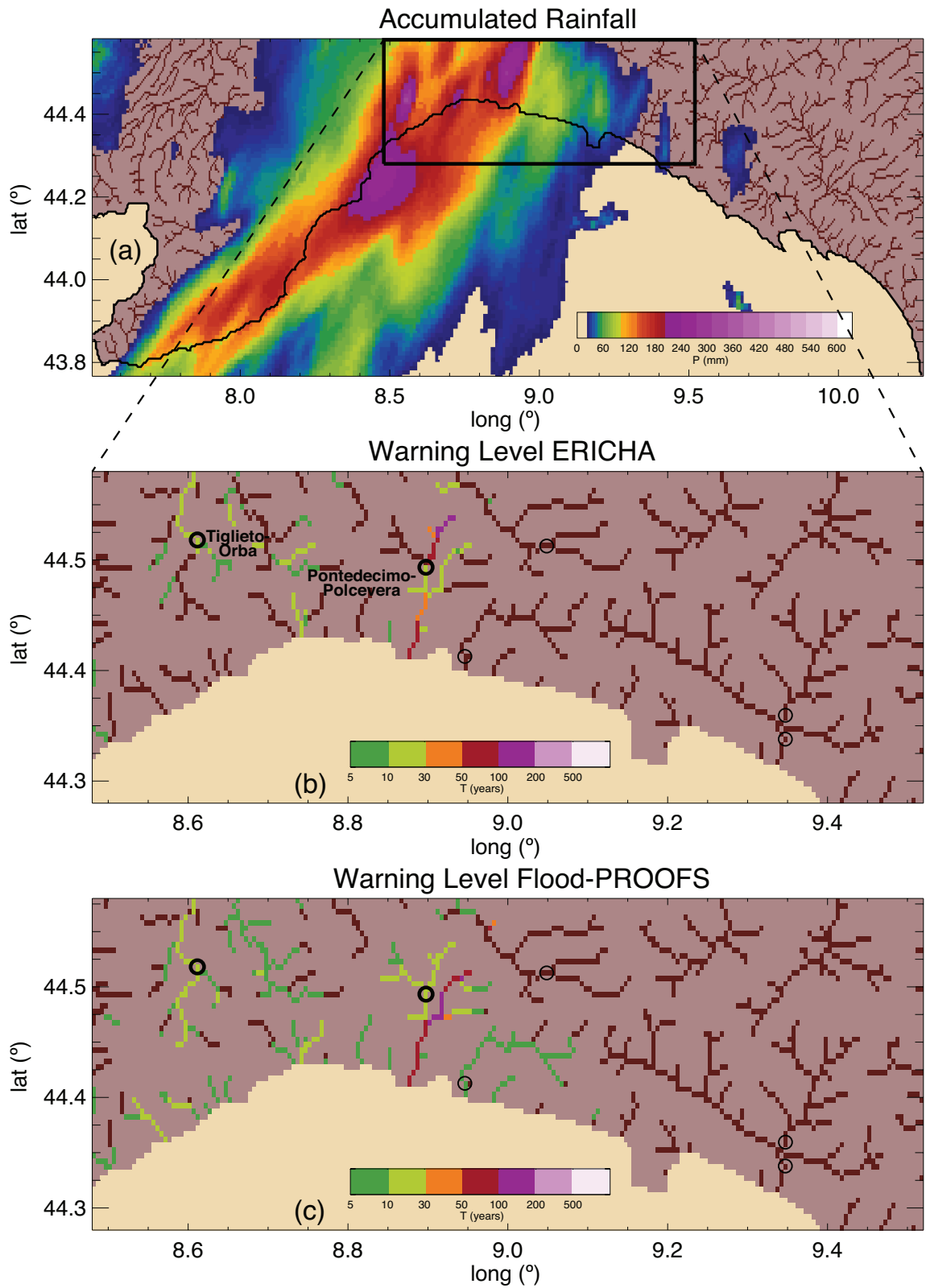

Figure 10. Rainfall accumulation for the event of 15 November from 14/10/2014 00:00 to 16/10/2014 00:00 (48-hour accumulation); b) Maximum return period obtained by ERICHA; c) Maximum return period obtained by Flood-PROOFS. Thicker circles

highlight the control points selected to show the temporal evolution in subsequent figures.

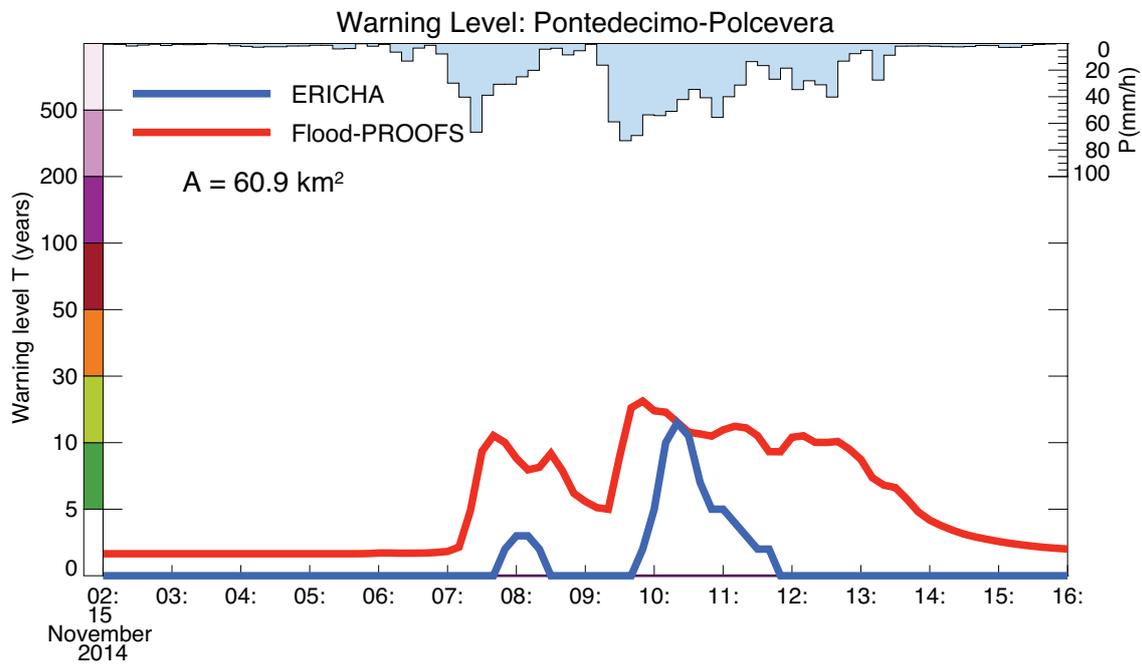

Figure 11. Comparison between ERICHA and Flood-PROOFS outputs for the event of 15 November 2014 at Pontedecimo control point (Polcevera river). Upstream catchment rainfall is included.

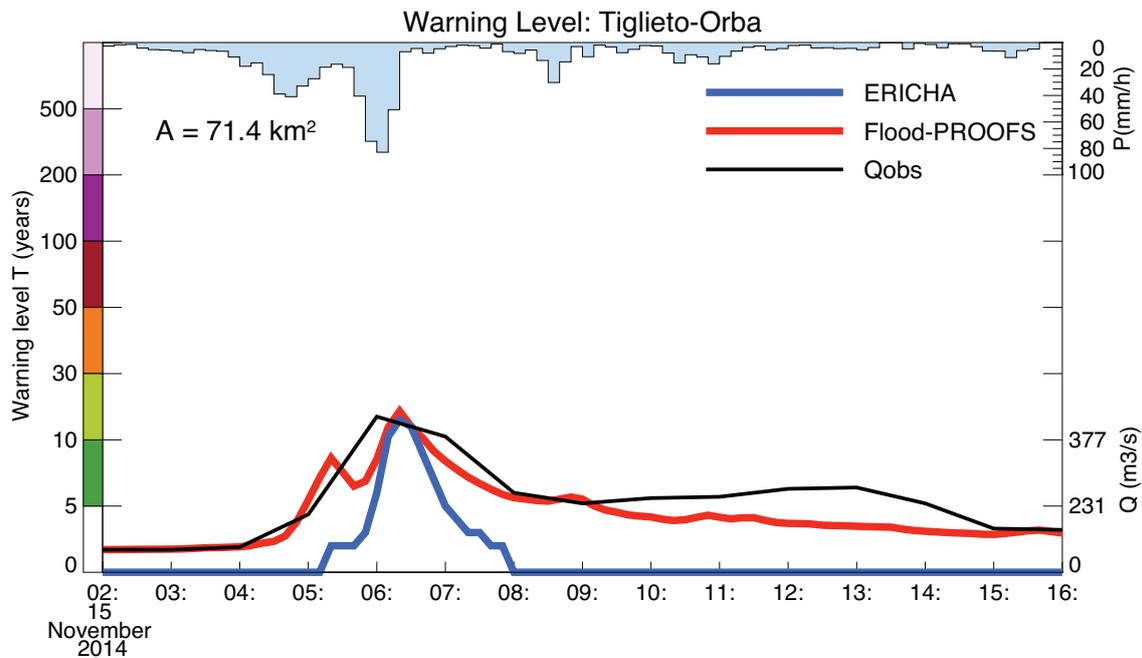

Figure 12. Comparison between ERICHA and Flood-PROOFS outputs for the event of 15 November 2014 at Tiglieto control point (Orba river). Warning level from observed flow is also included (threshold discharge values in the right axis). Upstream catchment rainfall is included.

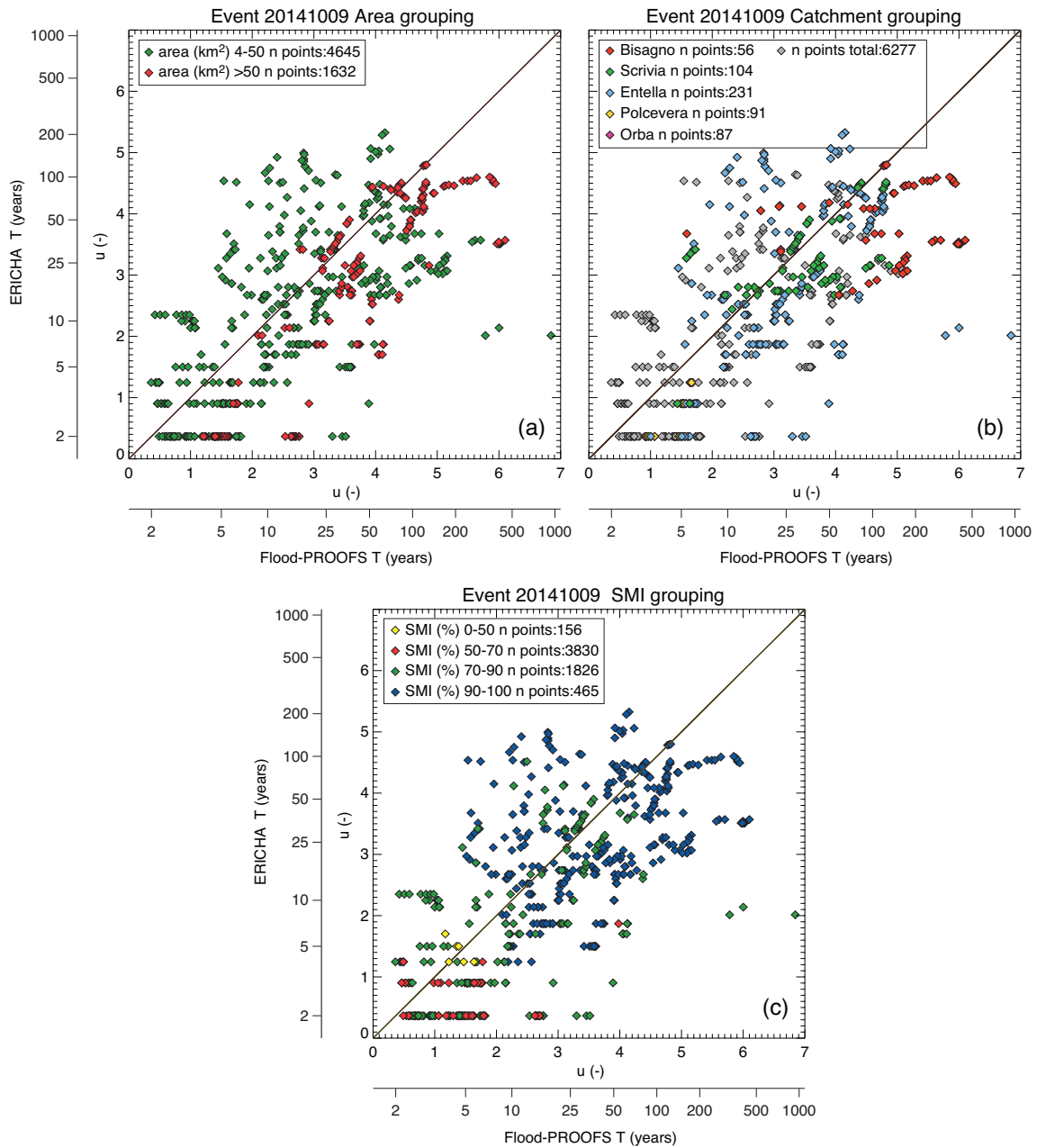

Figure 13. Scatter plot comparison between event maximum outputs from ERICHA and Flood-PROOFS at pixel scale for the 09 October 2014 event. Original return period values have been transformed to the Gumbel reduced variable (double logarithm). a)

Grouping is defined by point catchment area; b) Grouping is defined by catchment pertaining; c) Grouping is defined by upstream catchment soil moisture.

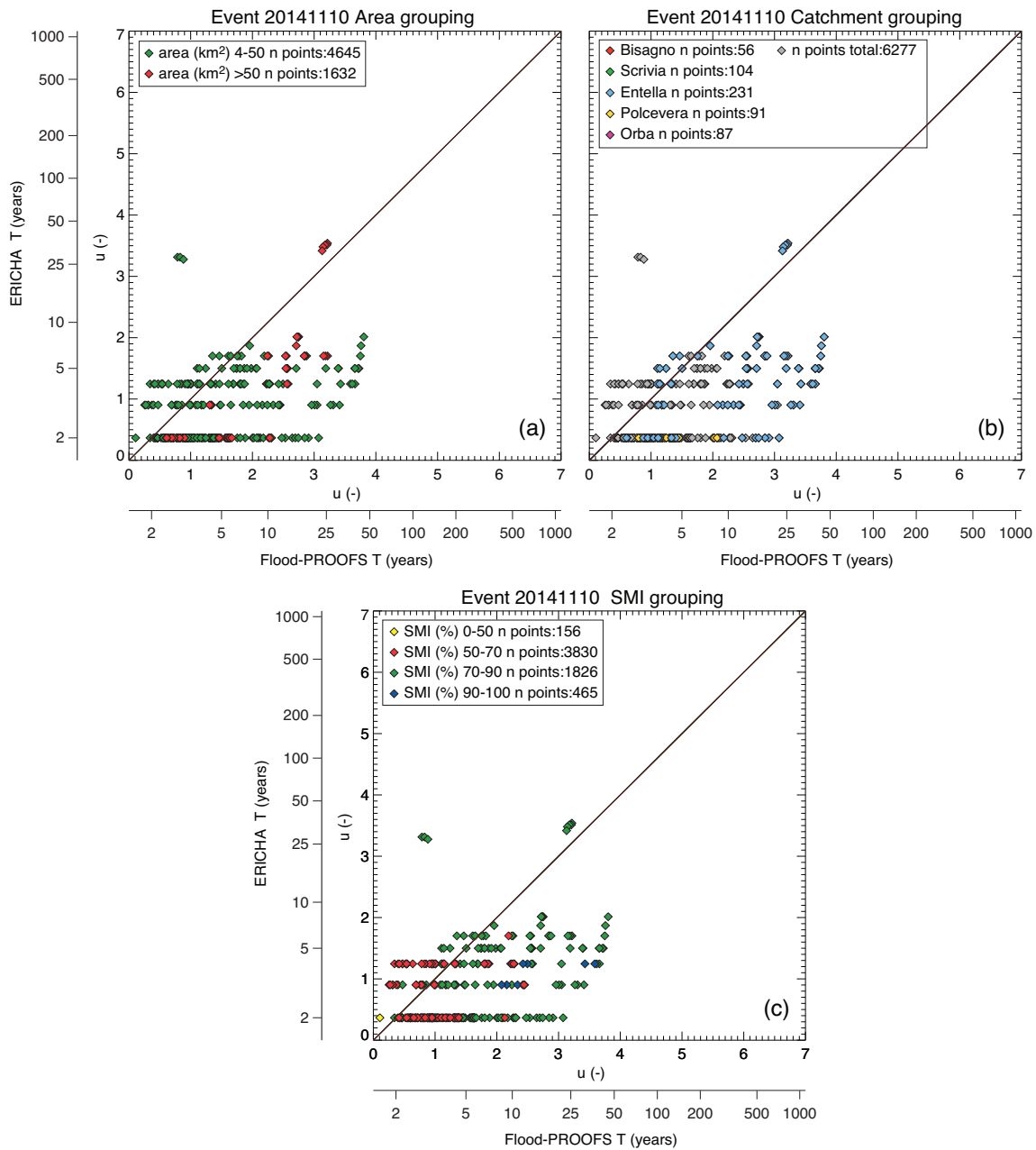

Figure 14. Scatter plot comparison between event maximum outputs from ERICHA and Flood-PROOFS at pixel scale for the 10 November 2014 event. Original return period values have been transformed to the Gumbel reduced variable (double logarithm). a) Grouping is defined by point catchment area; b) Grouping is defined by catchment pertaining; c) Grouping is defined by upstream catchment soil moisture.

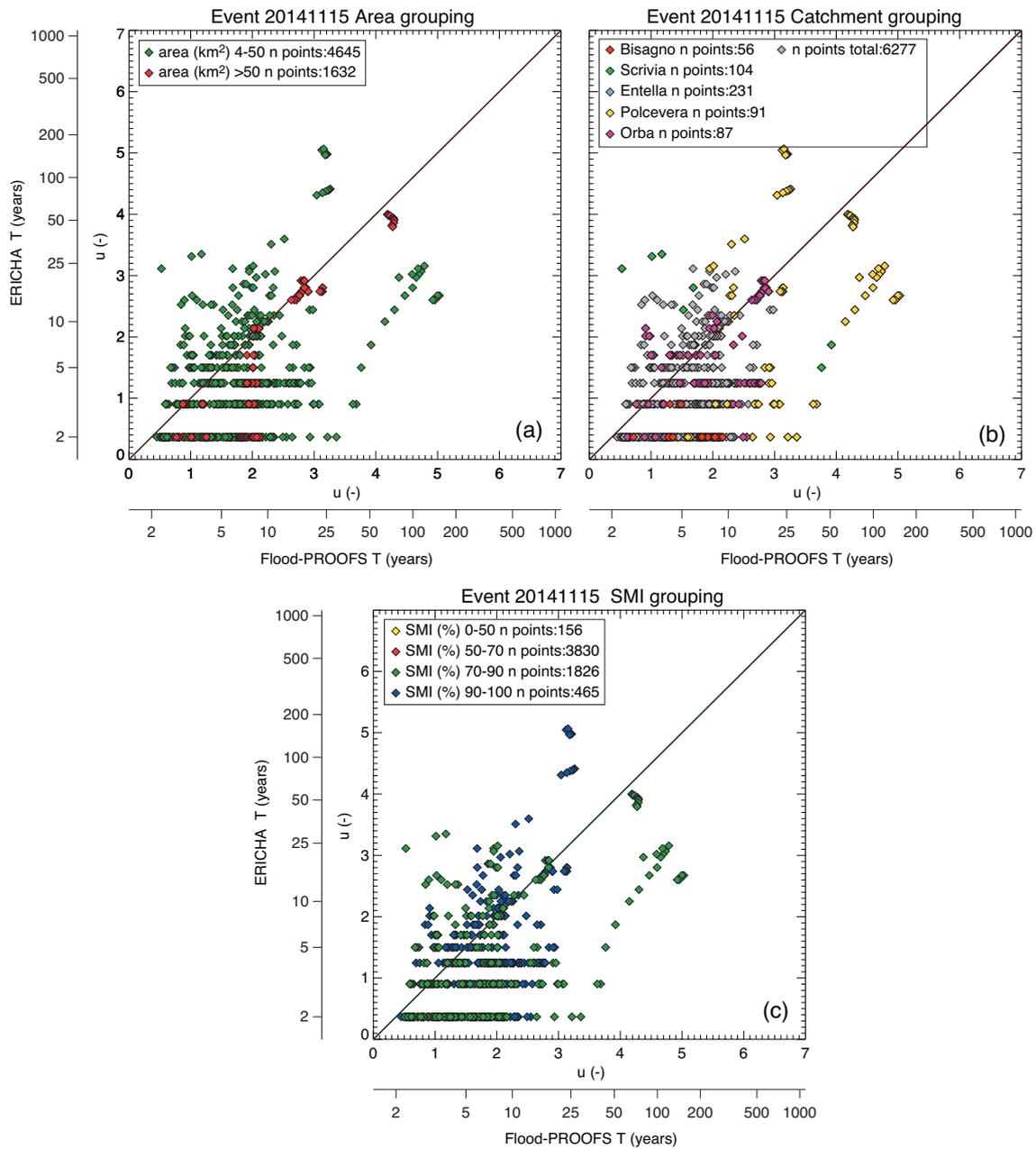

Figure 15. Scatter plot comparison between event maximum outputs from ERICHA and Flood-PROOFS at pixel scale for the 15 November 2014 event. Original return period values have been transformed to the Gumbel reduced variable (double logarithm). a) Grouping is defined by point catchment area; b) Grouping is defined by catchment pertaining; c) Grouping is defined by upstream catchment soil moisture.